\documentclass[]{emulateapj}

\begin{document}

\slugcomment{Accepted in ApJ}

\newcommand{\kms}{km s$^{-1}$}
\newcommand{\msun}{$M_{\odot}$}
\newcommand{\rsun}{$R_{\odot}$}
\newcommand{\vsini}{$v\sin{i}$}
\newcommand{\teff}{$T_{\rm eff}$}
\newcommand{\logg}{$\log{g}$}
\newcommand{\tarr}{$t_{\rm arr}$}
\newcommand{\mas}{mas yr$^{-1}$}

\title{ The Origin of HVS17, an Unbound Main Sequence B Star at 50 kpc }

\author{Warren R.\ Brown$^1$,
	Judith G.\ Cohen$^2$,
	Margaret J.\ Geller$^1$, and
	Scott J.\ Kenyon$^1$}

\affil{ $^1$Smithsonian Astrophysical Observatory, 60 Garden St, Cambridge, MA 02138\\
	$^2$Palomar Observatory, Mail Stop 249-17, California Institute of Technology, Pasadena, CA 91125 }

\email{wbrown@cfa.harvard.edu}
% jlc@astro.caltech.edu, mgeller@cfa.harvard.edu, skenyon@cfa.harvard.edu}

\shorttitle{ The Origin of HVS17 }
\shortauthors{Brown et al.}

\begin{abstract}
	We analyze Keck ESI spectroscopy of HVS17, a B-type star traveling with a
Galactic rest frame radial velocity of +445 \kms\ in the outer halo of the Milky
Way.  HVS17 has the projected rotation of a main sequence B star and is chemically
peculiar, with solar iron abundance and sub-solar alpha abundance.  Comparing
measured \teff\ and \logg\ with stellar evolution tracks implies that HVS17 is a
$3.91\pm0.09$~\msun, $153\pm9$~Myr old star at a Galactocentric distance of
$r=48.5\pm4.6$ kpc.  The time between its formation and ejection significantly
exceeds 10 Myr and thus is difficult to reconcile with any Galactic disk runaway
scenario involving massive stars.  The observations are consistent, on the other
hand, with a hypervelocity star ejection from the Galactic center.  We show that
{\it Gaia} proper motion measurements will easily discriminate between a disk and
Galactic center origin, thus allowing us to use HVS17 as a test particle to probe
the shape of the Milky Way's dark matter halo.

\end{abstract}


\section{INTRODUCTION}

	The origin of unbound stars in the Milky Way is linked to our 
understanding of physical processes in the Galaxy.  Unbound neutron stars, for 
example, are likely remnants of asymmetric core-collapse supernova explosions 
\citep[e.g.][]{arzoumanian02}.  Unbound white dwarfs, such as US 708 
\citep{hirsch05} and LP-400 \citep{kilic13}, are likely former binary companions 
of objects that exploded in double-detonation supernovae \citep{justham09, 
wang09, geier13}.  Unbound main sequence stars, stars that have not exploded, are 
a new probe of these issues.

	There are two models for explaining unbound main sequence stars: runaway
ejections from the Galactic disk, and hypervelocity star (HVS) ejections from the
Galactic center.  Traditional runaway B stars \citep[e.g.][]{humason47} are
either the former binary companions of systems disrupted by supernovae
\citep{blaauw61} or ejections in dynamical encounters with other stars
\citep{poveda67}.  \citet{heber08} first showed that an unbound runaway can
result from an extreme ejection from the outer disk in the direction of disk
rotation.  The physical size of main sequence stars places a natural speed limit
on runaway ejections, however, and so unbound runaways should be rare compared to
HVSs \citep[see][]{bromley09}.

	\citet{hills88} predicted that unbound ``hypervelocity stars'' are a natural 
consequence of a massive black hole (MBH).  There is overwhelming evidence for a 
$4\times10^6$~\msun\ MBH in the center of the Milky Way \citep{ghez08, gillessen09}.  
Theorists predict that 3-body interactions with this MBH will eject unbound HVSs at 
a rate of $\sim$$10^{-4}$ yr$^{-1}$ \citep{hills88, yu03, perets07, zhang10}.  The 
``S-stars'' presently observed in short-period, eccentric orbits around the MBH 
match expectations for being the former companions of ejected HVSs 
\citep{alexander04, ginsburg06, perets09c, zhang13, madigan13}.

	The first example of an unbound HVS is a short-lived B-type star traveling
at twice the Galactic escape velocity at a distance of $\simeq$100 kpc
\citep{brown05}.  After two other serendipitous HVS discoveries \citep{hirsch05,
edelmann05}, a targeted HVS survey by \citet{brown06, brown06b, brown07a, brown07b,
brown09a, brown12b} discovered 16 unbound B-type stars and a similar number of
possibly bound HVSs.  The existing observations -- the unbound velocities, the
observed number of HVSs, and the Galactic latitude distribution of HVSs -- support
the MBH ejection picture.  A handful of HVSs bright enough for echelle spectroscopy
are confirmed main sequence B stars at 50-100 kpc distances \citep{przybilla08,
przybilla08b, lopezmorales08, brown12c}.  These distances imply $\simeq$100 Myr
elapsed between the HVSs' formation and ejection, a timescale that is difficult to
reconcile with any runaway scenario involving supernovae or encounters with massive
stars that live for $\simeq$10 Myr, but consistent with the expected timescale for
dynamical interactions of stars with the central MBH \citep{brown12c}.  What remains
unclear, however, is whether all of the claimed HVSs are Galactic center ejections.

	A Galactic center origin for the HVSs links them with tidal disruption
events \citep{bromley12}.  HVSs and tidal disruption events both involve stars
formed in the central regions that are scattered into a MBH's ``loss cone.''
Observations imply a tidal disruption rate of $10^{-5}$ yr$^{-1}$ per galaxy
\citep[e.g.][]{vanvelzen11b}, however models are required to interpret the tidal
disruption light curves.  If HVSs are ejected from the Galactic center, then HVSs
provide a direct measure of the stellar interaction rate for a $4\times10^6$~\msun\
MBH.

	A Galactic center origin for the HVSs also makes them unique and important 
probes of the Milky Way's dark matter distribution \citep{gnedin05, yu07, wu08}.  
If HVSs are launched from $r=0$, then they integrate the Milky Way's gravitational 
potential as they travel out to 100 kpc distances.  Any deviation of the HVSs' 
trajectories from the Galactic center thus measures the Milky Way's mass 
distribution \citep{gnedin05}.  This measurement requires both proper motions 
\citep{brown10b} and accurate distances.  Proper motions and distances can also 
discriminate between an HVS origin from the Galactic center and a runaway origin 
from the outer disk \citep{heber08, tillich09, irrgang10, tillich11}.

	High resolution spectroscopy allows us to measure accurate stellar
parameters and thus an improved distance.  The discriminatory power of these
observations is important because evolved blue horizontal branch (BHB) stars can
have the same temperature and gravity as late B-type main sequence stars, but with
very different luminosities.  For example, a BHB star with the temperature and
gravity of HVS17 is 8 times less luminous \citep{dorman93, dotter07}, and thus is
located at a 2.8 times nearer distance, than a main sequence B star of the same
temperature and gravity.  This distinction matters to our flight time and proper
motion calculations, as well as to our understanding of the mass function of stars
encountering the MBH.  Fortunately, projected stellar rotation \vsini\ provides a
clean discriminant between evolved and main sequence late B-type stars.  Evolved BHB
stars have median \vsini\ $=9$~\kms; the most extreme BHB rotation known is 40~\kms\
\citep{behr03}.  Late B-type main sequence stars, on the other hand, have median
\vsini\ $=150$~\kms; the most extreme main sequence star rotation exceed 350~\kms\
\citep{abt02, huang06a}.

	Here, we present a study of SDSS J164156.39+472346.1, henceforth HVS17, a
newly discovered HVS \citep{brown12b} bright enough for high resolution spectroscopy
with the 10m Keck telescope.  HVS17 appears to be a chemically peculiar B star,
which means that diffusion processes have erased any constraint on stellar origin
provided by abundance.  Stellar rotation suffers from no such ambiguity, and on this
basis we conclude that HVS17 is a main sequence B star at a distance of 50 kpc.  We
investigate HVS17's origin on the basis of trajectory and flight time calculations.
If HVS17 is a runaway ejected from the Galactic disk, it requires a minimum ejection
of +415 \kms\ to reach its present location and velocity.  A Galactic center origin,
on the other hand, implies a proper motion that differs from the disk origin by
$\simeq$1 mas yr$^{-1}$, a difference easily measurable in the near future with {\it
Gaia}.

	In Section 2 we describe the observations and stellar atmosphere analysis.  
In Section 3 we discuss the nature and origin of HVS 17.  We conclude in Section 4.

\section{DATA}

\subsection{Observations}

	We observed HVS17 using the ESI spectrograph \citep{sheinis02} at the 10~m 
Keck~2 telescope.  We used the 0.5~arcsec slit to obtain a spectral resolution of 
$R\simeq$9,000; the spectral coverage is 3900--9300 \AA.  We collected seven 30 min 
exposures over the course of three nights 2012 April 26-28.

	We use the pipeline package {\sc makee}\footnote{{\sc~makee} is a 
spectroscopic reduction package developed by T.A.\ Barlow.  It is freely available 
from the Keck HIRES home page \url{www2.keck.hawaii.edu/inst/hires}.} to extract a 
one dimensional spectrum for each echelle order and calibrate the wavelength scale 
using arc spectra with Cu, Xe, Ne, Ar, and Hg lines.  Comparison of arc and sky 
lines show that the wavelength shift between the three nights is less than 0.6 
pixel (7 \kms).  Each exposure of HVS17 was individually processed, and the results 
summed.  Our total integration time of 3.5 hours achieves a $S/N$ ratio of 300 per 
spectral resolution element in the continuum at 4500~\AA.

	We also observed four B stars selected from \citet{abt02} that span a wide 
range in projected rotational velocity ($15<$~\vsini~$<240$~\kms).  The stars are 
HR5833, HR5834, HR6502, and HR6851.  We used the high S/N spectra of these four B 
stars to validate our stellar atmosphere analysis below.

\begin{figure}		% FIGURE 1:
 % \plotone{/pool/wbrown0/Bcand/Keck2/Dat/tplot.ps}
 \plotone{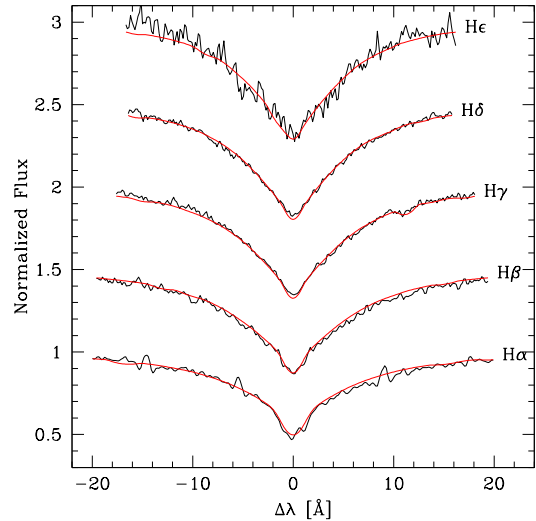}
 \caption{ \label{fig:tplot}
	Observed hydrogen Balmer lines compared to the best-fit model (smooth
lines).  The temperature- and surface gravity-sensitive lines give best-fit values
of \teff$=12350\pm290$ K and \logg$=3.80\pm0.086$.}
 \end{figure}

\subsection{Spectral Analysis}

	Our spectral analysis methodology is described in \citet{brown12c}.  In a 
sentence, we calculate synthetic spectra using the {\sc spectrum} package 
\citep{gray94} and ATLAS9 ODFNEW model atmosphere grids \citep{castelli04, 
castelli97}, normalize the continuum, calculate the $\chi^2$ of each synthetic model 
against the data, and then fit the resulting distribution of $\chi^2$ to derive the 
best-fitting parameters and uncertainties.

	The best-fit $+248.0\pm2.2$~\kms\ radial velocity is in good agreement with 
the $+246\pm9$~\kms\ radial velocity measured from medium-resolution spectroscopy at 
the MMT \citep{brown12b}; there is no evidence for velocity variability.  HVS17's 
heliocentric velocity corresponds to a minimum velocity of +445~\kms\ in the 
Galactic rest frame \citep[see][]{brown12b}.

	Next, we measure projected rotation using Mg {\sc ii} $\lambda$4481 \AA, the 
single strongest metal line in the spectrum.  The minimum $\chi^2$ is sensitive to 
Mg abundance, but insensitive to \teff\ or \logg.  Iterating with the best-fit 
parameters below, we find \vsini $=68.7\pm5.4$~\kms.  

	Given the observed \vsini, we measure effective temperature and surface 
gravity from the \teff- and \logg-sensitive hydrogen Balmer lines.  The best-fit 
values are \teff$=12,350\pm290$ K and \logg$=3.80\pm0.086$; Figure~\ref{fig:tplot} 
compares the best-fit model with the data.

\begin{figure}		% FIGURE 2:
 % \plotone{/pool/wbrown0/Bcand/Keck2/Dat/mplot.ps}
 \plotone{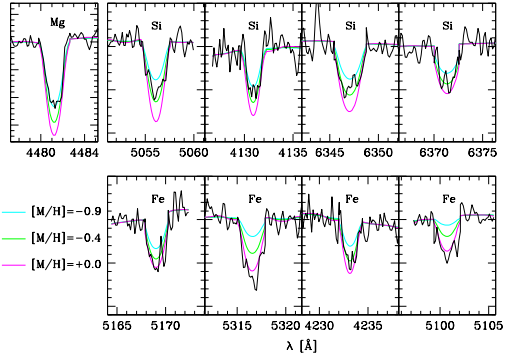}
 \caption{ \label{fig:mplot}
	The strongest metal lines in the HVS17 spectrum (black lines) compared to
models with [M/H]=-0.9 (cyan line), [M/H]=-0.4 (green line), and [M/H]=+0.0 (magenta
line) for the best-fit \teff, \logg, and \vsini.  Mg best matches [M/H]=-0.9, the Si 
lines on average match [M/H]=-0.4, and the Fe lines on average match [M/H]=+0.0.}
 \end{figure}

	Finally, we measure metal abundances by generating a list of all metal lines
with $>$10 m\AA\ equivalent widths, and then averaging the abundances measured for
all lines of a given element.  We exclude lines blended by another species.  
Figure~\ref{fig:mplot} plots the strongest metal lines in the spectrum compared with
some fiducial models.  Iron is the best-constrained element, with 48 unblended Fe
{\sc ii} lines and a weighted mean of [Fe/H]=$+0.06\pm0.22$.  Silicon is also
well-measured, with 9 unblended Si {\sc ii} lines and a weighted mean of
[Si/H]=$-0.39\pm0.18$.  C, Mg, S, and Ti have only a few lines and abundances
ranging $-0.9 < {\rm [M/H]} < -0.7$ dex.  Table~\ref{tab:param} summarizes all of
the measured parameters for HVS17.

\begin{figure}		% FIGURE 3
 % \plotone{/pool/wbrown0/Bcand/Padova/teff17.ps}
 \plotone{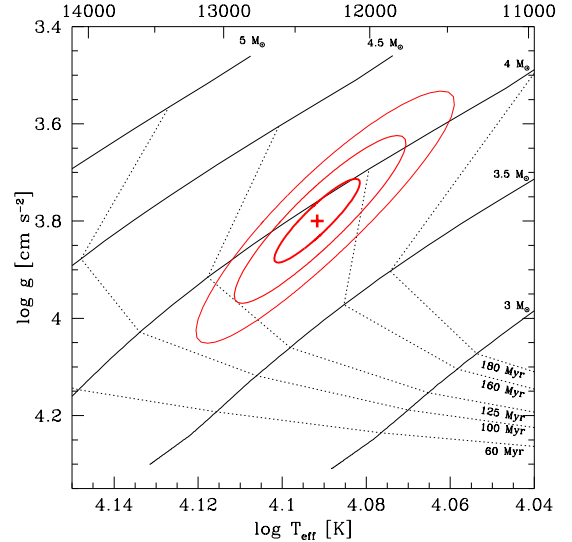}
 \caption{ \label{fig:teff}
	Measured 1-, 2- and 3-$\sigma$ confidence regions (ellipses) compared to
\citet{marigo08} solar metallicity main sequence tracks (straight solid lines).  
Isochrones (dotted lines) are plotted for the solar metallicity tracks.  We conclude
HVS17 is a $153\pm9$ Myr old $3.91\pm0.09$ \msun\ star. }
 \end{figure}

\section{DISCUSSION}

\subsection{Stellar Nature}

	HVS17 has the \teff\ and \logg\ of either a main sequence B star or a hot
BHB star.  Thus we turn to metallicity and rotation to determine its nature.  Main
sequence B stars presumably come from metal-enriched star formation regions, whereas
evolved BHB stars are normally found in old metal-poor environments like the stellar
halo.  Curiously, HVS17 has a solar iron abundance and sub-solar alpha abundance.  
Anomalous abundance patterns are seen in some hot BHB stars due to their shallow
surface convection zones \citep{michaud08}.  Yet hot BHB stars are among the slowest
rotators, with observed \vsini $\le7$ \kms\ \citep{behr03, behr03b}.  If HVS17 is a
hot BHB star, it is the fastest rotating hot BHB star by a factor of ten.  An
extreme BHB rotation might be explained if the star was spun-up and ejected by a
binary MBH \citep{lockmann08}, however there is presently no evidence for a binary
black hole in the Galactic center.

	The observed \vsini\ of HVS17 is consistent, on the other hand, with the
\vsini\ of late B-type main sequence stars \citep{abt02, huang06a}.  Chemically
peculiar A- and B-type stars are common, because of atomic diffusion processes in
the radiative atmospheres of these stars \citep{michaud70}.  Chemically peculiar
stars are also slower rotators on average than non-peculiar main sequence stars
\citep{smith96}.  The observed \vsini\ of HVS17, which is about 100 \kms\ below the
mean \vsini\ of single late B-type main sequence stars \citep{abt02, huang06a},
matches this expectation.  Slower-than-average rotation also fits the expectation
for 3-4 \msun\ HVS ejections from a single MBH, in which tidal synchronization of
the progenitor stellar binary yields \vsini$\sim$80 \kms\ \citep{hansen07}. In
either case, the observations are consistent with HVS17 being a main sequence star,
and a chemically peculiar B star like HVS7 \citep{przybilla08b}.

	Figure~\ref{fig:teff} compares the measured \teff\ and \logg\ with Padova
\citep{girardi02, girardi04, marigo08} solar metallicity (solid lines) main sequence
tracks.  Diffusion processes have erased any constraint provided by abundances if
HVS17 is chemically peculiar.  In other words, we do not know HVS17's interior
composition and cannot say whether it is consistent or inconsistent with expected
Galactic center abundance patterns.  For discussion, we will compare with
solar metallicity tracks because HVS17's iron abundance is solar.  The ellipses in
Figure~\ref{fig:teff} plot the measurement 1-, 2- and 3-$\sigma$ confidence regions.  
Interpolating these tracks indicates that HVS17 is a $3.91\pm0.09$ \msun\ star that
is $153\pm9$ Myr old.  The age uncertainty is relatively small because \teff\ and
\logg\ change rapidly with increasing age, as illustrated by the isochrones (dotted
lines) in Figure ~\ref{fig:teff}.  The absolute $g$-band magnitude
$M_g=-1.05\pm0.19$ places HVS17 at a heliocentric distance of $49.6\pm4.6$ kpc.  
Assuming the Sun is located 8~kpc from the Galactic center, HVS17's Galactocentric
distance is $r=48.5 \pm4.6$ kpc.

	To estimate systematic uncertainty we compare with \citet{bressan12} tracks
which use a different definition of solar metallicity.  These tracks yield a mass of
3.71 \msun\ and an age of 175 Myr, values which differ from \citet{marigo08} tracks
by twice our 1-$\sigma$ errorbars.  Choice of tracks clearly introduces a systematic
uncertainty:  lower metallicity tracks yield lower masses, lower luminosities (and
thus shorter distances and flight times), and increased ages.  Fortuitously, these
trends are in a direction that strengthen our conclusions below.

\begin{deluxetable}{llll}		% Stellar Parameters
\tablewidth{0pt}
\tablecaption{Stellar Parameters\label{tab:param}}
\tablecolumns{4}
\tablehead{ \colhead{Measured} & & \colhead{Derived\tablenotemark{a}} & }
	\startdata
\teff\ (K)	& $12350 \pm 290  $	& Mass (\msun)		& $3.91  \pm 0.09 $ \\
\logg\ (cgs)	& $ 3.80 \pm 0.086$	& Radius (\rsun)	& $4.12  \pm 0.23 $ \\
\vsini\ (\kms)	& $ 68.7 \pm 5.4  $	& Age (Myr)		& $153   \pm 9    $ \\
$\rm [C/H]$	& $-0.84 \pm 0.40 $	& $M_g$ (mag)		& $-1.05 \pm 0.19 $ \\
$\rm [Mg/H]$	& $-0.90 \pm 0.30 $	& $r_{GC}$ (kpc)	& $48.5  \pm 4.6  $ \\
$\rm [Si/H]$	& $-0.39 \pm 0.18 $	& $t_{GC}$ (Myr)	& $90    \pm 6    $ \\
$\rm [S/H]$	& $-0.70 \pm 0.35 $	& Age$-t_{GC}$ (Myr)	& $63    \pm 11   $ \\
$\rm [Ti/H]$	& $-0.90 \pm 0.30 $	& & \\
$\rm [Fe/H]$	& $+0.06 \pm 0.22 $	& & \\
$g_0$ (mag)	& $17.428\pm 0.015$	& & \\
$v_{\rm helio}$ (\kms) & $+248.0\pm 2.2$& & \\
	\enddata
\tablenotetext{a}{Derived quantities assume solar-metallicity (see text).}
 \end{deluxetable}

\subsection{Origin}

	Our observations paint the following picture:  HVS17 is a short-lived main 
sequence B star in the outer halo, traveling at an unbound radial velocity.  
Galactic escape velocity at $r=50$ kpc is approximately 400 \kms\ for a Milky 
Way halo mass of $1.6\times10^{12}$ \msun\ \citep{gnedin10}, and HVS17's radial 
velocity in the Galactic rest frame is +445 \kms.  Two models that can explain 
HVS17's origin are a hyper-runaway ejection from the Galactic disk and a 
hypervelocity ejection from the Galactic center.

	The mass and surface gravity of HVS17 yield an escape velocity from the
surface of the star of 602 \kms.  If we take this velocity as the speed limit on any
runaway origin, whether supernovae or dynamical ejection from 3- or 4-body stellar
encounters, then there is a finite part of the Galactic disk from which HVS17 can be
ejected.  We therefore use trajectory calculations to constrain HVS17's origin.  
Our approach is to start at the present position and radial velocity of HVS17 and
calculate backwards in time trajectories that cross the Galactic disk.

\begin{figure}		% FIGURE 4
 % \plotone{/pool/wbrown0/Bcand/Keck2/Dat/xy.ps}
 \plotone{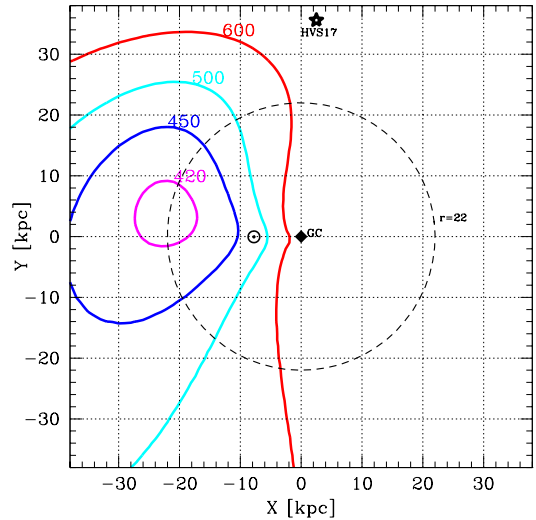}
 \caption{ \label{fig:xy}
	Galactic plane ejection velocities (contours) required to place HVS17 at 
its present position (star) and radial velocity.  The Sun is at X=$-8$ kpc and 
Galactic rotation is in the clockwise direction.  The minimum ejection velocity 
is 415 \kms\ at a distance of $r=22$ kpc (dashed circle); the escape velocity 
from the surface of HVS17 is 600 \kms.}
 \end{figure}

	Figure \ref{fig:xy} plots the ejection velocity from the Galactic disk
required to place HVS17 at its present position and radial velocity.  Our
calculation assumes the Galactic potential model of \citet{kenyon08} and a 250 \kms\
circular velocity \citep{reid09, mcmillan10}.  The lowest possible ejection velocity
is 415 \kms\ at a distance of $r=22$ kpc from the Galactic center; ejections from
inside the solar circle require velocities in excess of 500 \kms.  For context,
theoretical models predict that less than 1\% of runaway ejections, whether from
supernovae \citep{portegies00} or dynamical ejections \citep{perets12}, have
$v_{ej}>200$ \kms.  Extreme runaway ejection velocities require massive stars
\citep{heber08, przybilla08c, gvaramadze09b}.

	For the well-defined region in Figure \ref{fig:xy} with $v_{ej}<420$ 
\kms, the flight time from the disk to the location of HVS17 is $96\pm4$ Myr.  
Thus the time between when HVS17 formed and when it was ejected is $57\pm10$ Myr 
in the disk runaway scenario.  This time between formation and ejection is 
difficult to reconcile with any runaway ejection involving massive stars that 
live for only $\sim$10 Myr.  Ejection during the first 10 Myr of HVS17's lifetime 
is formally ruled out at the 5-$\sigma$ level, and by a larger amount if HVS17 is 
a lower mass or lower metallicity (and thus older) star.

	We will henceforth refer to the time between HVS17's formation and ejection
as its ``arrival time'' \citep{brown12c}.  For a Galactic center origin, there is no
upper limit on arrival time.  The massive black hole is always there, and on-going
star formation \citep[e.g.][]{lu09} provides a constant supply of new stars.  
Theorists predict that dynamical interactions and orbital evolution within a
triaxial potential will cause stars formed in the central region to ``fill'' the
black hole's loss cone with timescales of 100~Myr to 1~Gyr \citep{yu03, merritt04,
wang04, perets07}.  Arrival time thus clearly distinguishes between the central
black hole and disk runaway ejection processes.

	The flight time from the Galactic center to the location of HVS17 is
$90\pm6$ Myr.  The uncertainty comes from propagating the distance and radial
velocity errors through the trajectory calculation.  The arrival time for the
Galactic center scenario is thus $63\pm11$ Myr.  This timescale is consistent with
the timescale for stars to form in the Galactic center region and scatter into the
black hole's loss cone.  Three-body interactions with the massive black hole
naturally provide unbound ejection velocities \citep{hills88, yu03}, and at a rate
that is 100$\times$ larger than the ejection rate of unbound disk runaways
\citep{brown09a, perets12}.

\begin{figure}		% FIGURE 5
 % \plotone{/pool/wbrown0/Bcand/Keck2/Dat/pm.ps}
 \plotone{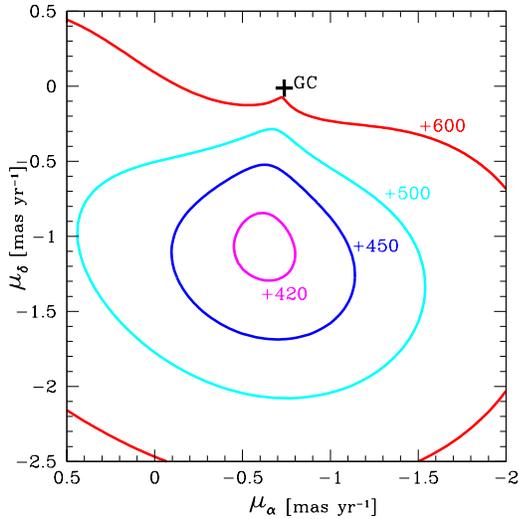}
 \caption{ \label{fig:pm}
	Predicted proper motion for a Galactic center origin (GC) or Galactic disk 
origins with the indicated ejection velocities (contours).}
 \end{figure}

\subsection{Proper Motion Prediction}

	It appears that HVS17 comes from the Galactic center based on its unbound
velocity and $\gg$10 Myr arrival time, however a more direct test will soon be
possible:  proper motion.  Although the expected proper motion of HVS17 is
$\lesssim$1 \mas\ because of the star's distance, the direction differs for 
Galactic center and Galactic disk origins.

	We plot proper motions corresponding to different HVS17 trajectories in
Figure \ref{fig:pm}.  For a Galactic center ejection, we predict that HVS17 has
$(\mu_{\alpha}, \mu_{\delta}) = (-0.75, 0.0)$ \mas.  For a Galactic disk ejection
with $v_{ej}<420$ \kms, HVS17 has $(-0.8 < \mu_{\alpha} < -0.5, -1.3 < \mu_{\delta}
< -0.9)$ \mas.  This disk origin differs from the Galactic center origin by
$\simeq$1 \mas.  Higher disk ejection velocities allow for a broader range of proper
motion, but nearly all physically possible disk ejections require trajectories with
more southerly proper motions than trajectories from the Galactic center.  This
difference in proper motion should be easily measured by {\it Gaia}, which will
achieve $\pm0.045$ \mas\ precision 
({\tiny \url{http://www.rssd.esa.int/index.php?page=Science\_Performance\&project=GAIA}})
for this star.

\section{CONCLUSION}

	We present Keck ESI spectroscopy of HVS17, a late-B type star traveling with
a minimum Galactic rest frame velocity of +445~\kms.  HVS17 has a projected rotation
of \vsini$=68.7\pm5.4$~\kms\ and thus is a main sequence B star.  The star appears
chemically peculiar with solar iron abundance and sub-solar alpha abundance.  
Diffusion processes have thus erased any constraint provided by abundances.  
Comparing measured \teff\ and \logg\ with solar metallicity stellar evolution tracks
implies that HVS17 is a $3.91\pm0.09$~\msun, $153\pm9$~Myr old star at a distance of
$r=48.5\pm4.6$ kpc.  Sub-solar metallicity tracks systematically increase HVS17's 
inferred age.

	We establish HVS17's origin using velocity and ``arrival time,'' the time
between its formation and subsequent ejection.  A disk runaway origin suffers a
fatal lifetime problem:  the required $>$415 \kms\ ejection velocities require
massive stars that live for only $\sim$10 Myr.  For the part of the Galactic disk
from which ejection velocities are less than the escape velocity from the surface of
HVS17, arrival times significantly exceed 10 Myr.  The central black hole origin, on
the other hand, allows for any arrival time.  The central black hole is also
expected to eject unbound 3-4 \msun\ stars at a rate 100$\times$ larger than disk
runaway scenarios \citep{brown09a, perets12}.  We conclude that HVS17 is likely a
HVS ejected by the MBH in the Galactic center.

	Future proper motion measurements will directly answer the question of
origin.  We predict that trajectories from the Galactic disk differ systematically
by $\simeq1$ \mas\ compared to the trajectory from the Galactic center.  This
difference is easily measurable with {\it Gaia}.  If HVS17 is indeed ejected from
the Galactic center, its proper motion, coupled with our measurement of nature and
distance, will one day allow us to use it as a test particle for mapping the Milky
Way's dark matter distribution.

\acknowledgements

	This work was supported in part by the Smithsonian Institution.  
J.~Cohen acknowledges partial support from AST--0908139.  This research makes use 
of NASA's Astrophysics Data System Bibliographic Services. We are grateful to the 
many people who have worked to make the Keck Telescopes and their instruments a 
reality, and who operate and maintain these observatories. The authors wish to 
extend special thanks to those of Hawaiian ancestry on whose sacred mountain we 
are privileged to be guests.  Without their generous hospitality, none of the 
observations presented herein would have been possible.

\noindent {\it Facility:}  Keck:II (ESI)

	% REFERENCES 
% \clearpage
% \bibliographystyle{/home/wbrown/lib/apj} \bibliography{/home/wbrown/text/RefHS}

\end{document}